\documentclass[twocolumn,pre,twoside,showpacs,preprintnumbers,floatfix]{revtex4}
\usepackage{graphicx}% Include figure files
\usepackage{dcolumn}% Align table columns on decimal point
\usepackage{bm}% bold math
\begin{document}
\title{Depletion potential in colloidal mixtures of hard spheres 
and platelets}
\author{L. Harnau and S. Dietrich}
\affiliation{
         Max-Planck-Institut f\"ur Metallforschung,  
         Heisenbergstr.\ 3, D-70569 Stuttgart, Germany, 
         \\
         and Institut f\"ur Theoretische und Angewandte Physik, 
         Universit\"at Stuttgart, 
         Pfaffenwaldring 57, 
         D-70569 Stuttgart, Germany
	 }
\date{\today}
\begin{abstract}
The depletion potential between two hard spheres in a solvent of thin 
hard disclike platelets is investigated by using either the Derjaguin 
approximation or density functional theory. Particular attention is paid to 
the density dependence of the depletion potential. A second-order virial 
approximation is applied, which yields nearly exact results for the bulk 
properties of the hard-platelet fluid at densities two times smaller 
than the density of the isotropic fluid at isotropic-nematic phase 
coexistence. As the platelet density increases, the attractive primary 
minimum of the depletion potential deepens and an additional small 
repulsive barrier at larger sphere separations develops. Upon 
decreasing the ratio of the radius of the spheres and the 
platelets, the primary minimum diminishes and the position of the small 
repulsive barrier shifts to smaller values of the sphere separation.
\end{abstract}

\pacs{61.20.-p, 61.20.Gy, 82.70.Dd}
\maketitle
\section{Introduction} 
 
Depletion interactions between big colloidal particles induced 
by smaller particles, which can be either the solvent particles 
or a colloidal component of its own, are of significant current
research interest because of the importance of these effective 
interactions in various colloidal processes. For example, flocculation 
of colloids can be driven by the addition of non-adsorbing polymers via 
the depletion mechanism \cite{loui:01}. Whereas experimental and 
theoretical studies have focussed on binary hard sphere fluids as 
well as on colloidal mixtures of hard spheres and hard rods or 
polymers, less attention has been paid to hard platelets acting as 
depletants, despite the great importance of colloidal platelets 
such as blood-platelets and clay minerals in both biomedicine and 
geophysics. Very recently a colloidal mixture of silica 
spheres and silica 
coated gibbsite platelets has been stabilized 
for the first time \cite{over:03}, and the depletion potential 
due to the presence of thin hard platelets has been derived 
theoretically for non-interacting platelets corresponding to 
the limit of infinite dilution \cite{over:03,piec:00}. 
It has been found that the Derjaguin approximation for the depletion 
potential yields accurate results for non-interacting platelets
provided the ratio of the radius of the spheres and the platelets
is large \cite{over:03}.

In this paper we focus on the depletion interaction induced 
by thin hard platelets, taking into account the steric interactions
between the platelets in terms of a second-order virial approximation. 
On the basis of our recent theoretical studies on 
fluids of thin hard platelets near hard walls \cite{harn:02a,harn:02c}, 
we expect that excluded volume interactions between the platelets
influence the depletion interaction already at rather low platelet 
densities due to their cumbrous shape as compared with spherical 
or rodlike depletants. Taking excluded volume interactions into account
is particularly interesting because correlation effects may cause 
repulsive features of depletion forces which are important in the 
context of colloidal stability \cite{mao:97}. 
In the present paper we use density functional theory (Sec. II) 
to study the depletion potential between two hard spheres 
induced by thin hard platelets (Sec. III). Particularly, we 
compare the results with the ones obtained for non-interacting 
platelets.

\section{Density Functional Theory}
We consider an inhomogeneous fluid consisting of thin platelets of radius 
$R_p$ in a container of volume $V$. The platelets are taken to be hard discs 
without additional attractive or repulsive interactions. The number density 
of the centers of mass of the platelets at a point ${\bf r}$ with an 
orientation $\omega=(\theta,\phi)$ of the normal of the platelets  is 
denoted by $\rho({\bf r},\omega)$. The equilibrium 
density profile of the inhomogeneous liquid under the influence 
of an external potential $V({\bf r},\omega)$ minimizes 
the grand potential functional
\begin{eqnarray} \label{eq1}
\Omega[\rho({\bf r},\omega)]\!\!\!&=&\!\!\!
\int dr^3\,d\omega\,\rho({\bf r},\omega)
\left[k_BT\left(\ln[4\pi\Lambda^3\rho({\bf r},\omega)]-1\right)\right.\nonumber
\\ &&-\left. \mu+ V({\bf r},\omega)\right]+
F_{ex}[\rho({\bf r},\omega)]\,,\nonumber
\\
\end{eqnarray}
where $\Lambda$ is the thermal de Broglie wavelength and $\mu$ is the 
chemical potential. The free energy functional 
$F_{ex}[\rho({\bf r},\omega)]$ in excess of the ideal gas 
contribution has not been taken into 
account in previous studies on the depletion force due to  
platelets \cite{piec:00,over:03}. We express the excess free energy functional
as an integral over all possible configurations of two platelets 
\begin{eqnarray} 
F_{ex}[\rho({\bf r},\omega)]&=&-\frac{k_BT}{2}
\int dr^3_1\,d\omega_1
\,dr^3_2\,d\omega_2\,
\rho({\bf r}_1,\omega_1)\nonumber
\\&&\times f_{pp}({\bf r}_{12},\omega_1,\omega_2)
\rho({\bf r}_2,\omega_2)\,,\label{eq2}
\end{eqnarray}
where ${\bf r}_{12}={\bf r}_1-{\bf r}_2$ and 
$f_{pp}({\bf r}_{12},\omega_1,\omega_2)$
is the Mayer function of the interaction potential between two platelets. 
The Mayer function equals $-1$ if the platelets 
overlap and is zero otherwise. Explicit expressions of the Mayer 
function for thin platelets are documented in 
Refs.~\cite{harn:02a} and \cite{eppe:84}.

For the homogeneous and isotropic bulk fluid the grand potential 
functional [Eq.~(\ref{eq1})] reduces to 
\begin{eqnarray} \label{eq3}
\frac{\Omega_b}{V}=\rho_b\left[k_BT(\ln[\lambda^3\rho_b]-1)-\mu\right]+
\frac{\pi^2}{2}R_p^3\rho_b^2k_BT\,,
\end{eqnarray}
where $\rho_b=V^{-1}\int dr^3\,\int d\omega\, \rho({\bf r},\omega)$
is the total particle number density. The equation of state derived from 
the grand potential [Eq.~(\ref{eq3})] takes the following form:
\begin{equation} \label{eq4}
p^\star_b=\rho_b^\star\left(1+\frac{\pi^2}{2} \rho_b^\star\right)
\end{equation}
with $\rho_b^\star=\rho_bR_p^3$ and $p_b^\star=p_bR_p^3/(k_BT)$. The same 
equation without the second term in parenthesis holds for the ideal gas 
limit (i.e., non-interacting platelets). With increasing particle number 
density the ideal gas equation of state on one side and the second-term virial 
series [Eq.~(\ref{eq4})] as well as computer simulation data 
\cite{eppe:84,dijk:97,bate:99} on the other side deviate. Thus for 
$\rho_b^\star=0.1$ the osmotic pressure $p_b^\star=0.15$, 
as calculated from Eq.~(\ref{eq4}), agrees exactly with simulation data, 
while the ideal gas equation of state underestimates the osmotic pressure 
by a factor of $1.5$. The comparison of the calculated equation of state with 
computer simulation data exhibits that the two-term series in Eq.~(\ref{eq4}) 
is a good approximation for $\rho_b^\star\lesssim 0.2$, whereas the ideal 
gas model may be used for very low particle number densities
$\rho_b^\star\lesssim 0.04$. For a discussion of higher order virial terms 
for fluids consisting of hard platelets we refer to 
Refs.~\cite{harn:02a} and \cite{onsa:49,harn:01,harn:02b}. In the present study 
we restrict our attention to particle number densities $\rho_b^\star \le 0.2$
for which the second-order virial approximation is appropriate
and the platelet fluid is in the isotropic phase. For comparison, the 
isotropic-nematic phase transition is first order with coexistence densities 
$\rho_{bI}R_p^3=0.46$ and $\rho_{bN}R_p^3=0.5$ according to a computer 
simulation \cite{bate:99}.

\section{The platelet-induced depletion potential between two spheres}
\begin{figure}[t]
\begin{center}
\vspace*{-2cm}
\hspace*{-2cm}
\includegraphics[width=6cm]{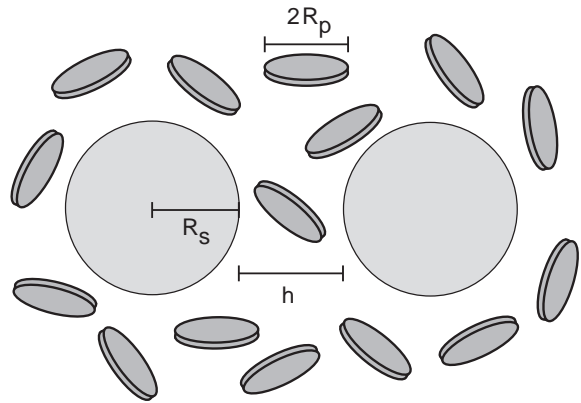}
\vspace*{0.5cm}
\caption{The system under consideration consists of two hard spheres of radius 
$R_s$ immersed in a solvent of hard platelets of radius $R_p$. The separation 
between the surfaces of the spheres is denoted by $h$. Only the projection 
of the spheres on the plane of the figure is shown.}
\label{fig1}
\end{center}
\end{figure}

The results of the preceding section show that intermolecular interactions 
between platelets increase the osmotic pressure of the bulk fluid already 
at low particle densities. Now we study the influence of intermolecular 
interactions on the depletion potential between two hard spheres 
of radius $R_s$ immersed in a fluid of hard platelets of radius $R_p$. 
The depletion potential $W(h)$ is the free energy difference between 
the configurations of two big spheres at fixed distance $h$ immersed 
in the solvent and at macroscopic separation $h=\infty$ 
(see Fig.~\ref{fig1}).

\subsection{The Derjaguin approximation}
The depletion potential $W(h)$ between two hard spheres at close distance
due to the presence of small platelets ($R_p \ll R_s$) can be calculated 
from the finite size contribution of the grand potential function 
$\omega(h')$ of the platelet fluid confined between two parallel hard walls 
at distance $h'$ using the Derjaguin approximation \cite{derj:34}
\begin{equation} \label{eq5}
W_{Derj}(h)=\pi R_s\int_h^\infty dh'\,\omega(h')\,,
\end{equation}
where $h$ is the separation between the surfaces of the spheres. (For the 
subtle issue of the range of validity of the Derjaguin approximation 
see Refs.~\cite{hend:02} and \cite{oett:03}.)

\begin{figure}[t]
\begin{center}
\vspace*{-1.0cm}
\includegraphics[width=9cm]{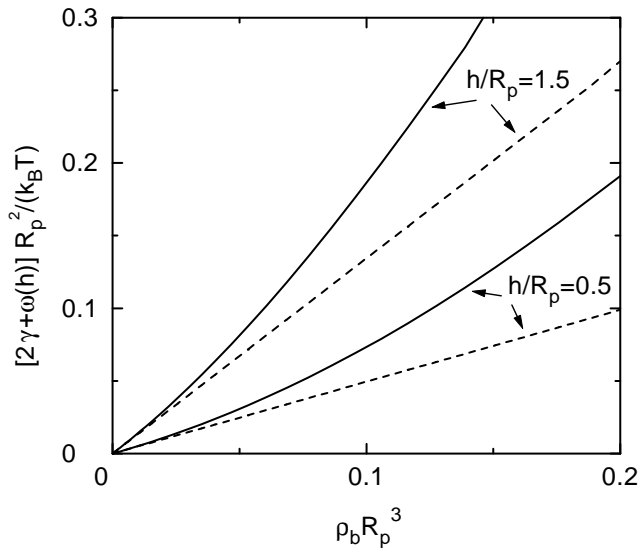}
\vspace*{-4.5cm}
\caption{The surface and finite size contribution $2\gamma+\omega(h)$ to the grand 
potential as obtained from Eqs.~(\ref{eq1}), (\ref{eq2}), and (\ref{eq6}) 
[solid lines] of a fluid consisting of thin hard platelets of radius $R_p$ confined 
in a slit of width $h$ and in contact with an isotropic bulk reservoir at density 
$\rho_b$. The dashed lines represent the corresponding results for an ideal gas 
of platelets [see Eq.~(\ref{eq7})]. The width of the slit increases from bottom to 
top: $h/R_p=0.5, 1.5$.}
\label{fig2}
\end{center}
\end{figure}

\begin{figure}[t]
\begin{center}
\vspace*{-0.2cm}
\includegraphics[width=9cm]{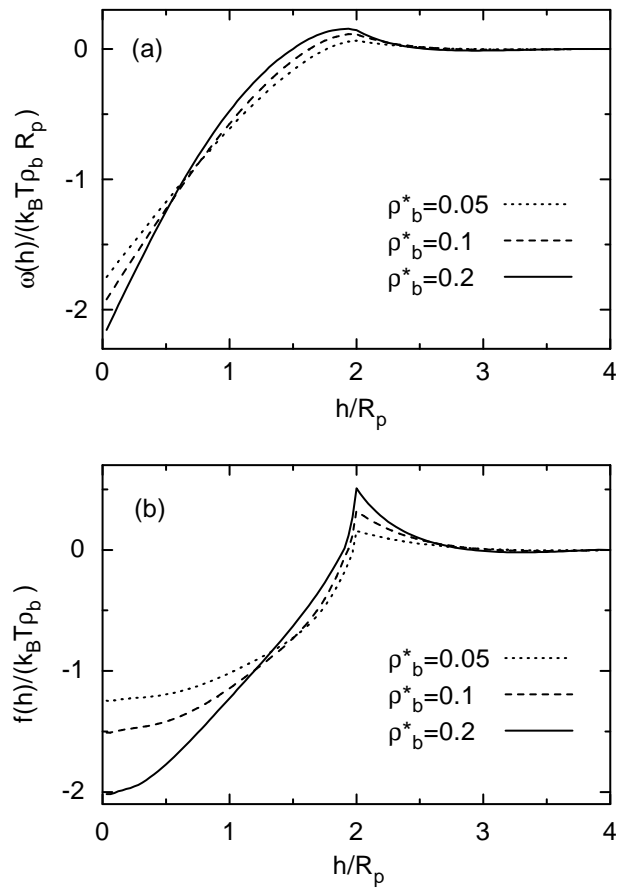}
\vspace*{-0.7cm}
\caption{(a) The finite size contribution $\omega(h)$ to the grand potential
as obtained from Eqs.~(\ref{eq1}), (\ref{eq2}), and (\ref{eq6}) 
of a fluid consisting of thin hard platelets of radius $R_p$ confined 
in a slit of width $h$ for three values of the density of 
the isotropic bulk reservoir: 
$\rho^\star_b=\rho_b R_p^3=0.05$ (dotted line);
$\rho^\star_b=0.1$ (dashed line); 
$\rho^\star_b=0.2$ (solid line). 
(b) The solvation force per unit area $f(h)=-d \omega(h)/dh$ of the same fluid 
[with the same line code as in (a)] as a function of $h$.}
\label{fig3}
\end{center}
\end{figure}

\begin{figure}[t]
\begin{center}
\vspace{-1.2cm}
\includegraphics[width=9cm]{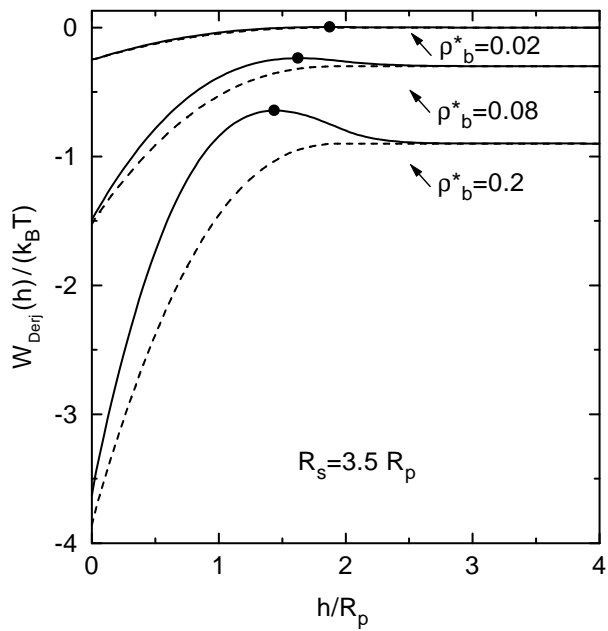}
\vspace{-3.3cm}
\caption{Depletion potential $W_{Derj}(h)$ between two hard spheres of 
radius $R_s=3.5 R_p$ due to the presence of thin hard platelets of radius 
$R_p$ as obtained from the Derjaguin approximation [Eq.~(\ref{eq5})]  
for various bulk densities $\rho_b$ of the platelets: 
$\rho_b^\star=\rho_b R_p^3=0.02$ (top curves);
$\rho_b^\star=0.08$ (middle curves);
$\rho_b^\star=0.2$  (bottom curves). The solid lines 
(with dots marking the maxima) represent the calculations for interacting 
platelets and the dashed lines denote the results for non-interacting platelets. 
For reasons of clarity, the lower four curves are shifted down by -0.3 $k_BT$ 
and -0.9 $k_BT$, respectively. The dashed lines are zero for $h/R_p\ge 2$. 
Although the interactions have only minor influence on the depth 
of the primary minimum, the depletion potential becomes significantly less 
attractive for increasing densities $\rho_p$. This weakens the platelet 
induced flocculation of a solution of big spheres.}
\label{fig4}
\end{center}
\end{figure}

We first consider a hard-platelet fluid confined by two parallel hard 
walls at $z=0$ and $z=h$, and calculate the surface and finite size
contributions to the grand potential defined via 
\begin{eqnarray} \label{eq6}
\Omega[\rho(z,\theta,\phi)]&=&V\omega_b+2A\gamma+A\omega(h)\,,
\end{eqnarray}
where $A$ is the area of a single surface, $\omega_b$ is the bulk 
grand canonical potential density, and $V$ is defined as the volume 
of the container with its surface given by the position of the rim 
of the particles at closest approach so that $V=Ah$. $\gamma$ is the 
wall--liquid surface tension in the absence of the second wall
and $\omega(h)$ is the finite size contribution. Figure \ref{fig2} 
displays the calculated surface and finite size contributions to the 
grand potential together with the results for non-interacting 
platelets \cite{piec:00,over:03}
\begin{equation} \label{eq7}
\frac{\left[2\gamma+\omega(h)\right]_{ideal}}{k_BT}\!\!=\!\!\left\{
\begin{array}{ll}
\rho_bR_p\left(\arcsin\left(\frac{h}{2R_p}\right)\right.
\\
\left.+\frac{h}{2R_p}\sqrt{1-\left(\frac{h}{2R_p}\right)^2}
\right),&\!\! 0\le h\le 2R_p\\
\frac{\pi}{2}\rho_bR_p=
\frac{2\gamma_{ideal}}{k_BT},& \!\!h\ge 2R_p\,.
\end{array}
\right.
\end{equation}
As is apparent from Fig.~\ref{fig2} the steric interaction between the 
platelets increases the surface contributions with increasing density. 
Within our numerical precision, we found that an accurate evaluation 
of the wall--liquid surface tension $\gamma$ could be achieved for a 
fixed wall separation $h=4R_p$ at all considered densities 
$\rho_bR_p^3\le 0.2$. For much higher densities, a larger value of $h$ 
might be required because of the wetting of the wall--isotropic 
liquid interface by a nematic film of diverging thickness \cite{harn:02c}.
On the other hand the wall--liquid surface tension 
$\gamma_{ideal}/(k_BT)=\rho_b R_p \pi/4$ for non-interacting platelets follows 
from Eq.~(\ref{eq7}) for a wall separation $h=2R_p$. For a detailed
discussion of the surface tension and the excess coverage as well as 
the density and orientational order parameter profiles of fluids 
consisting of thin hard platelets near a single hard wall we refer to 
Ref.~[4].
The results for the finite-size contribution $\omega(h)$ are shown 
in Fig.~\ref{fig3} (a). As function of $h$ the finite size contribution 
corresponds to the solvation free energy for the immersed two plates 
acting as the confining walls for the fluid and, by construction, 
$\omega(0)=-2\gamma$ and $\omega(\infty)=0$. Upon increasing 
the platelet density, the attractive minimum of $\omega(h)$ at $h=0$ 
deepens and a maximum at larger values of $h$ develops.
The corresponding solvation force per unit area $f(h)=-d \omega(h)/d h$ 
is attractive for small slit widths $h$ as is shown in Fig.~\ref{fig3} 
(b). Upon increasing the platelet density, the cusp of the solvation force 
at $h=2R_p$ sharpens. For comparison we note that the maximum at 
the cusp is more pronounced for the confined platelet fluid than for a 
corresponding rod fluid \cite{mao:97} due to the relatively larger 
steric interactions between platelets as compared with those between rods. 
Moreover $f(h)$ is a convex function for slit widths smaller than two 
times the radius of the platelets, while the solvation force in a solvent 
of hard rods is a concave function for slit widths smaller than the length 
of the rods \cite{mao:97}.
Figure \ref{fig4} displays the depletion potential together 
with the results for non-interacting platelets which can be calculated 
analytically from Eqs.~(\ref{eq5}) and (\ref{eq7}) 
\cite{over:03,piec:00}:
\begin{equation} \label{eq7a}
\frac{W_{Derj}^{(ideal)}}{k_BT}\!=\!\left\{
\begin{array}{ll}
-\pi\rho_bR_p^2 R_s\left(\frac{h}{R_p}\arcsin\left(\frac{h}{2R_p}\right)\right.
\nonumber\\\left.
+\frac{4}{3}\sqrt{1-\left(\frac{h}{2R_p}\right)^2}
\left(1+\frac{h^2}{8R_p^2}\right)-\frac{\pi h}{2R_p}\right)\,,
\\ \hspace{1cm} 0\le h\le 2R_p
\\0, \hspace{0.69cm} h\ge 2R_p\,.
\end{array}
\right.
\end{equation}
The depletion potential due to the presence 
of interacting platelets exhibits a  small barrier at larger 
sphere separations $h$ in addition to the primary minimum at $h=0$.
With increasing platelet density the depletion potential deepens 
and the position of the maximum shifts to smaller values of $h$. 
The small repulsive barrier will have minor effects on 
kinetic stabilization, although the repulsive features might still be 
measurable. 
For example, our numerical calculations exhibit a maximum 
barrier height of $0.25\,\, k_BT$ relative to zero at a density of 
$\rho_b R_p^3=0.2$ in a system of size ratio $R_s/R_p=3.5$ corresponding 
to the aforementioned mixture of silica spheres and gibbsite platelets 
\cite{over:03}. In order to examine the influence of steric 
interactions between platelets on thermodynamic properties of 
sphere-platelet mixtures, we treat the depletion potential as a 
perturbation to the hard-sphere potential. The first-order approximation 
in this thermodynamic perturbation approach for the Helmholtz free 
energy per sphere is  
\begin{eqnarray} \label{eq7b}
f(\rho_s,\rho_b)\!\!\!&=&\!\!\!f^{(hs)}(\rho_s)\nonumber
\\&&+2\pi\rho_s\!\!\!\int\limits_{2R_s}^\infty dr\,r^2 W(r-2R_s) 
g^{(hs)}(r,\rho_s),
\end{eqnarray}
where $g^{(hs)}(r,\rho_s)$ is the radial distribution function of 
the pure sphere fluid and $f^{(hs)}(\rho_s)$ is the Helmholtz free energy 
per sphere for the same homogeneous fluid of density $\rho_s$. In the limit 
$R_s\gg R_p$ the radial distribution function $g^{(hs)}(r,\rho_s)$ 
is almost constant over the range of integration where $W(r-2R_s)\neq 0$
and we can approximate it by its constant contact value 
$g^{(hs)}(2R_s,\rho_s)$ \cite{hans:86}.
Using Eq.~(\ref{eq7a}) resulting the integral in Eq.~(\ref{eq7b}) can be
evaluated analytically for non-interacting platelets:
\begin{eqnarray} \label{eq7c}
A_{Derj}^{(ideal)}\!\!\!&\equiv&\!\!\! 2\pi\rho_s\int\limits_{2R_s}^\infty dr\,r^2 
\frac{W_{Derj}^{(ideal)}(r-2R_s)}{k_BT}\nonumber
\\&=&\!\!\!-2\pi^2\rho_bR_p^3\rho_sR_s^3
\left(\pi+\frac{64}{45}\frac{R_p}{R_s}+\frac{\pi}{12}\frac{R_p^2}{R_s^2}
\right)\!\!.
\end{eqnarray}
The integrated strength of the depletion potential $A_{Derj}^{(ideal)}$ 
is negative, reflecting the fact the depletion potential is always
attractive (Fig.~\ref{fig4}). A numerical calculation of the corresponding
quantity $A_{Derj}$ for interacting platelets exhibits that the steric 
interacting between the platelets weakens the integrated strength of the 
depletion potential by 22 \% at a bulk density $\rho_b R_p^3=0.2$.
Hence the influence of steric interactions between platelets might be 
quite visible for phase equilibria. For example, the thermodynamic onset of 
flocculation of colloidal spheres induced by the depletion effect will be 
reduced due to platelet interactions.

\subsection{Density functional approach}
A general approach for calculating the depletion potential is 
based on a density functional theory (DFT) for a mixture of hard 
spheres and the particles acting as depletants \cite{roth:00,roth:03}.
This approach avoids the Derjaguin approximation. Translating the 
particle insertion idea developed in Refs.~\cite{roth:00} and 
\cite{roth:03} to the present system leads to the following expression 
for the depletion potential between two spheres 
due to the presence of hard platelets:
\begin{equation} \label{eq8}
-\frac{W({\bf r})}{k_BT}\!\!=\!\!\frac{1}{4\pi}\int dr^3_1\,
d\omega\,
\left[\rho({\bf r}_1,\omega)-
\rho(\infty,\omega)\right]
f_{sp}({\bf r}-{\bf r}_1,\omega)\,
\end{equation}
where $f_{sp}({\bf r}-{\bf r}_1,\omega)$ is the 
Mayer function of the interaction potential between a sphere 
and a platelet. The Mayer function equals $-1$ if the particles 
intersect or touch each other and is zero otherwise. 
$\rho({\bf r}_1,\omega)$ is the density profile
of platelets in the external potential of a {\it single} fixed hard sphere
located at the origin of the coordinate system 
and $\rho(\infty,\omega)$ is the corresponding density profile of the 
bulk fluid. We emphasize that the density profile entering Eq.~(\ref{eq8}) 
depends only on equilibrium properties of the depletant fluid in the 
{\it absence} of the second hard sphere to be inserted at position ${\bf r}$. 
This observation simplifies the calculation of $W({\bf r})$ 
considerably, because the symmetry of the density profile is 
determined solely by the symmetry of the external potential of 
a single sphere fixed at the origin of the coordinate system. 

Apart from  possible surface freezing at high densities, 
non-uniformities of the density depend only on the radial distance
$r=|{\bf r}|$, so that $\rho({\bf r},\omega)=
\rho(r,\omega)$. Hence, calculating density 
profiles before insertion of the second hard sphere is much easier 
than after insertion, when the presence of the second sphere leads 
to a more complex spatial variation of the densities. 
A detailed discussion of Eq.~(\ref{eq8}) and its application to 
the analogous case of non-interacting hard rods acting as depletants 
is given in Ref.~\cite{roth:03}. 

For an ideal gas of platelets in contact with 
a fixed hard sphere the density profile reduces to 
$\rho(r_1,\omega)-
\rho(\infty,\omega)=
f_{sp}(r_1,\omega)$ so that the integral in 
Eq.~(\ref{eq8}) has a purely geometrical meaning and measures
the excluded volume of a platelet confined between two hard 
spheres located at the origin of the coordinate system and 
at position {\bf r}, respectively. 

In order to take intermolecular interactions between the 
platelets into account we first calculate numerically
the density profile $\rho(r,\omega)$ 
of platelets in an external potential of one fixed hard 
sphere of radius $R_s$. Thereafter the integral in Eq.~(\ref{eq8})
is evaluated by inserting this density profile. To our knowledge, 
this technique has not been used before for {\it interacting} 
non-spherical colloids acting as depletants. 

\begin{figure}[t]
\begin{center}
\vspace*{-1.2cm}
\includegraphics[width=8.8cm]{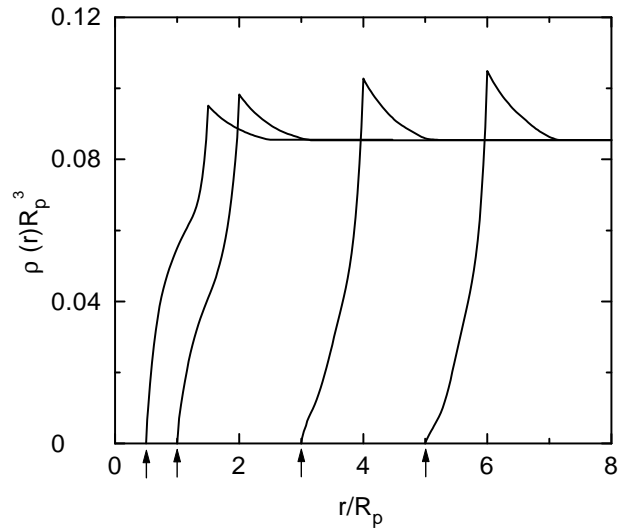}
\vspace*{-4.8cm}
\caption{Orientationally averaged density profile $\rho(r)$ as obtained from
Eq.~(\ref{eq9}) for hard platelets of radius $R_p$ in contact with a single hard 
sphere of radius $R_s$ located at $r=0$. The radius of the sphere increases 
from left to right: $R_s/R_p=0.5, 1, 3, 5$. The arrows mark the location of 
the surface of the sphere at $r=R_s$. Since the platelets are arbitrarily thin 
their density is nonzero for $r>R_S$. All curves exhibit a cusp at $r=R_s+R_p$ 
followed by a decay towards the bulk density $\rho_bR^3=0.085$, which is 
essentially reached at $r=R_s+2R_p$. When the center of a platelet is located 
less than $R_p$ from the sphere surface, there are fewer possible orientations 
available to the platelet.}
\label{fig5}
\end{center}
\end{figure}

\begin{figure}[th!]
\begin{center}
\includegraphics[width=7.3cm]{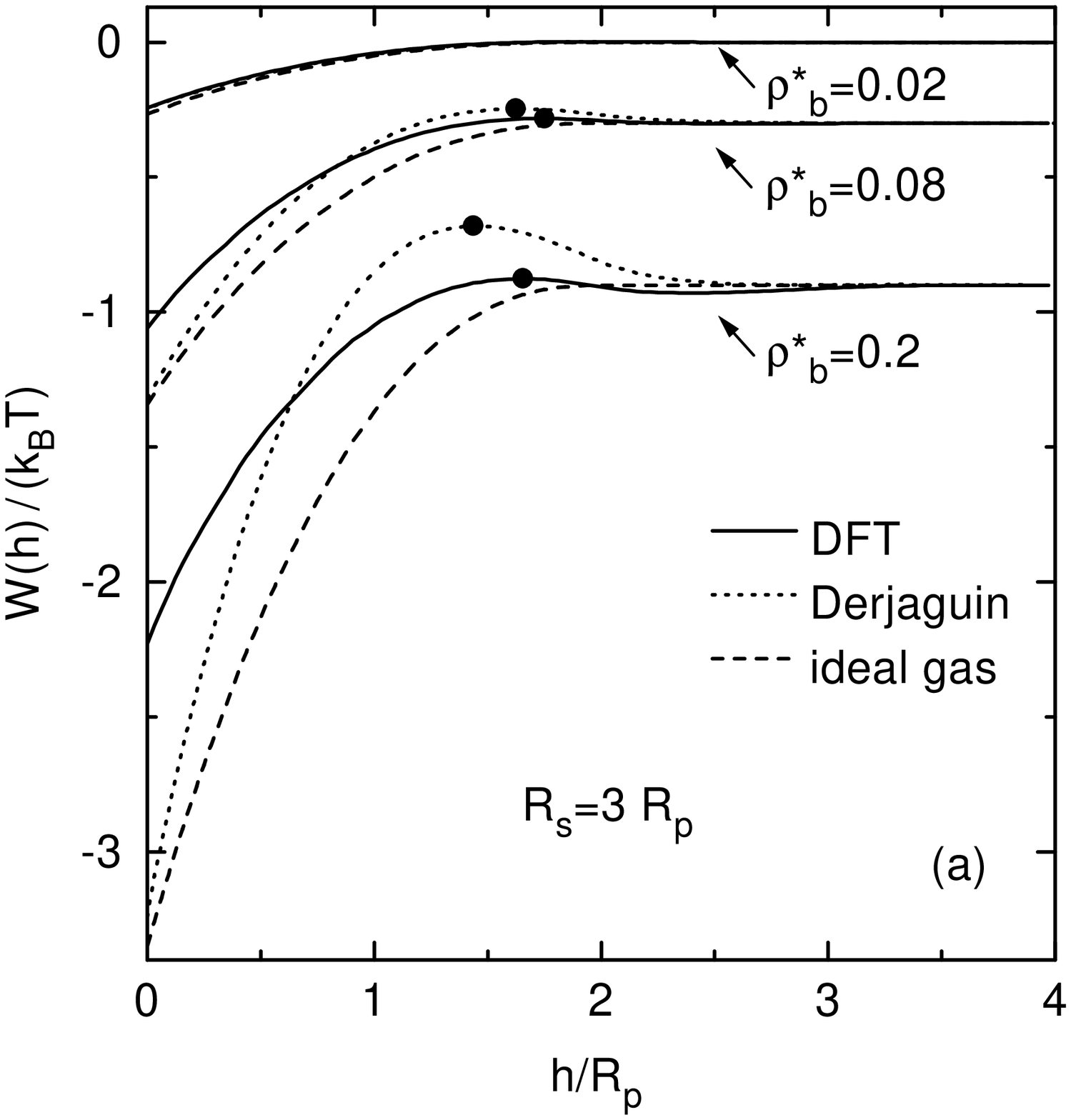}\\
\vspace{0.75cm}
\includegraphics[width=7.3cm]{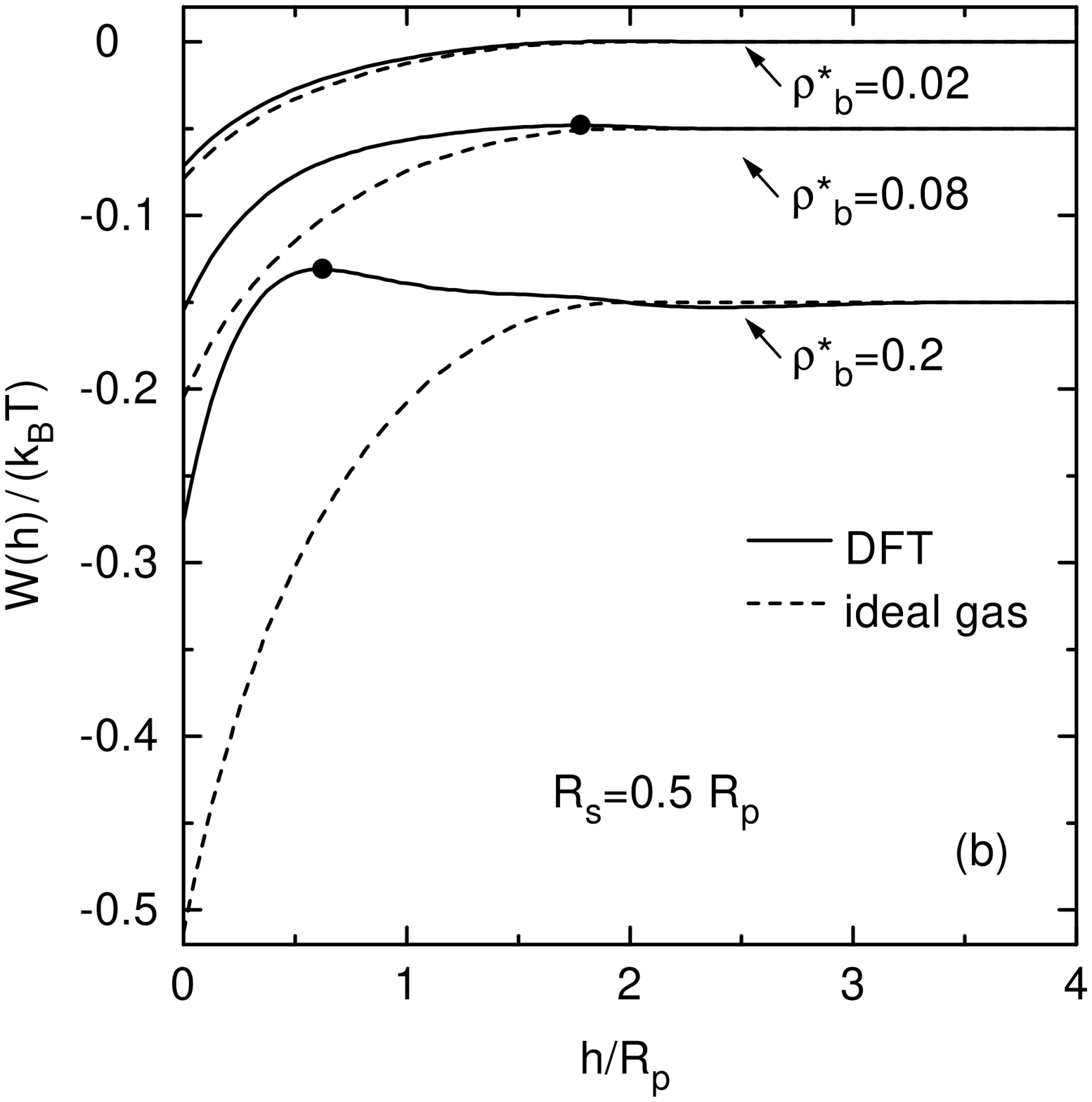}
\caption{Depletion potential $W(h)$ between two hard spheres of 
radius $R_s=3R_p$ in (a) and $R_s=0.5R_p$ in (b) due to the presence 
of thin hard platelets of radius $R_p$ as obtained from Eq.~(\ref{eq8}). 
The bulk densities of the platelets are: 
$\rho_b^\star=\rho_b R_p^3=0.02$ (top curves);
$\rho_b^\star=0.08$ (middle curves);
$\rho_b^\star=0.2$  (bottom curves). 
Here $h=r-2R_s$ is the separation between the surfaces of the spheres. 
The solid and dashed lines 
represent the calculations for interacting (DFT) and non-interacting platelets
(ideal gas), respectively. In addition the depletion potential as obtained from 
the Derjaguin approximation [Eq.~(\ref{eq5})] for interacting platelets
is displayed by dotted lines in (a). For the smallest bulk density the solid 
and dotted line nearly coincide in (a). Since in (b) the spheres are only half 
as big as the platelets, in this case the Derjaguin approximation is unsuitable 
and therefore not shown. Only for sufficiently large densities a maximum 
denoted by a dot occurs. For reasons of clarity, the lower sets of curves are 
shifted down by -0.3 $k_BT$ and -0.9 $k_BT$, respectively in (a) and by 
-0.05 $k_BT$ and -0.15 $k_BT$, respectively in (b).}
\label{fig6}
\end{center}
\end{figure}

The orientational averaged density profile 
\begin{equation} \label{eq9}
\rho(r)=\int d\omega\,
\rho(r,\omega)
\end{equation}
of the platelet fluid in contact with one fixed hard sphere 
is shown in Fig.~\ref{fig5} for various radii $R_s$ of the sphere.
Upon increasing $r\ge R_s$ from the surface of the sphere
the averaged number density increases and exhibits a cusp 
at $r=R_s+R_p$ where platelets with their normal perpendicular
to the radial direction touch the surface of the sphere with the rim.  
The maximum at the cusp is about 25\% 
above the bulk value $\rho_b R_p^3=0.085$ for a size ratio
$R_s/R_p=5$ and is less pronounced for smaller size ratios. 
The averaged density close to the surface of the sphere is 
larger for a small sphere than for a big one.
Figure \ref{fig6} displays the calculated 
depletion potential for two size ratios $R_s/R_p$ as a 
function of the separation between the surfaces of the 
spheres $h=r-2 R_s$. With decreasing size ratio the range 
and the depth of the primary minimum shrinks and the position 
of the small repulsive barrier observed at higher densities 
shifts to smaller values of $h$. Moreover, the height of the 
repulsive barrier decreases upon decreasing the size ratio. 
These results are due to the fact that the number of platelets 
contributing to the depletion potential decreases as the 
ratio of the radius of the spheres and the platelets becomes 
smaller at a fixed bulk density (see Fig~\ref{fig5}).

In agreement with a recent theoretical study \cite{over:03}
based on an evaluation of the excluded volume of a single 
platelet confined between two hard spheres,
we find that the Derjaguin approximation for the depletion 
potential in the presence of non-interacting platelets yields 
accurate results for large size ratios $R_s/R_p>1$. However, 
there are substantial deviations at higher densities as
can be seen from Fig.~\ref{fig6} (a). The absolute value 
$|W(h=0)|$ of the DFT solution at contact is smaller than 
the one obtained from the Derjaguin approximation, and the 
repulsive barrier is less pronounced. Hence, for a size 
ratio $R_s/R_p=3$ and a platelet density $\rho_bR_p^3=0.2$ 
the Derjaguin approximation overestimates the depth of the 
depletion potential at contact by 37 \% and the height of 
the repulsive barrier by 23 \%. Increasing the size ratio 
$R_s/R_p>3$ does not lead to a significant improvement of 
the Derjaguin approximation as compared to the results shown 
in Fig.~\ref{fig6} (a). For $R_s/R_p<1$ and higher densities 
$\rho_b$ the DFT results deviate strongly from the predictions
for non-interacting platelets [see $\rho_b^\star=\rho_b R_p^3=0.2$ 
in Fig.~\ref{fig6} (b)].
\section{Summary}
We have applied a density functional theory to fluids consisting 
of thin hard platelets confined between two hard spheres 
(Fig.~\ref{fig1}). Within the framework of a second-order 
virial approximation of the excess free energy functional, 
the depletion potential between the two spheres due 
to the presence of the platelets is determined 
numerically and compared with the corresponding 
results for non-interacting platelets. The main conclusions
which emerge from our study are as follows.

(1) Figure \ref{fig2} demonstrate that steric interactions 
between thin platelets of radius $R_p$ confined between two 
parallel hard walls increase the sum of the surface and 
finite size contribution to the grand potential significantly 
already at rather low platelet densities $\rho_bR_p^3 \gtrsim 0.025$.

(2) As function of the slit width $h$ the finite size contribution
to the grand potential of a slap of platelets exhibits a minimum at 
$h=0$ [Fig.~\ref{fig3} (a)]. A maximum at larger values of $h$ is 
found for higher platelet densities. The corresponding solvation force 
is attractive for small slit widths and exhibits a cusp at $h=2R_p$ 
[Fig.~\ref{fig3} (b)].

(3) The depletion potential between two spheres as calculated 
from the Derjaguin approximation exhibits an attractive primary 
minimum at contact which deepens upon increasing the platelet density. 
Moreover, a small repulsive barrier at larger sphere 
separations develops with increasing density (Fig.~\ref{fig4}).
We find that the depletion barrier relative to zero is typically less 
than the thermal energy $k_BT$, and therefore unlikely to significantly 
alter the kinetics of aggregation of the hard spheres at platelet 
densities smaller than two times the density of the isotropic 
phase at bulk isotropic-nematic coexistence. Nonetheless, with 
increasing platelet density the integrated strength of the 
effective interaction between the spheres becomes significantly
weaker and thus reduces the thermodynamic onset of flocculation.

(4) The orientational averaged density profile of a platelet fluid 
in contact with a single fixed hard sphere decreases towards the surface 
of the sphere because the range of accessible orientations is reduced 
when the particle approaches the sphere. It exhibits a cusp at the position
where the platelets lose contact with the surface of the sphere
(Fig.~\ref{fig5}). The maximum at the cusp decreases and the averaged 
density close to the surface increases as the ratio of the radius of 
the sphere and the platelets becomes smaller.

(5) With decreasing ratio of the radius of the spheres and 
the platelets, the primary minimum at contact and the small repulsive 
barrier of the depletion potential diminish and the position 
of repulsive barrier shifts to smaller values of the separation 
of the surfaces of the spheres (Fig.~\ref{fig6}). From our numerical 
results based on a density functional theory for a mixture of spheres 
and platelets we found that the Derjaguin approximation is valid for 
large size ratio and very small platelet density, but there are 
substantial deviations from the density functional results
at higher densities even for large size ratios [Fig.~\ref{fig6} (a)].
These deviations increase with increasing platelet density.
For small size ratios and high platelet densities the ideal gas 
approximation for the platelets becomes unsuitable [Fig.~\ref{fig6} (b), 
$\rho_b^\star=0.2$].

\end{document}